\documentclass[aps,prl,twocolumn,superscriptaddress, showpacs, showkeys]{revtex4}
\usepackage[dvips]{graphicx}

\usepackage{dcolumn}
\usepackage{bm}

\begin{document}


\title{Persistence of Covalent Bonding in Liquid Silicon Probed by Inelastic X-ray Scattering}

\author{J. T. Okada}
\affiliation{Institute of Space and Astronautical Science, Japan Aerospace Exploration Agency, Tsukuba, Ibaraki 305-8505, Japan}
\author{P. H.-L. Sit}
\affiliation{Department of Chemistry, Princeton University, Princeton, New Jersey 08544, USA}
\author{Y. Watanabe}
\affiliation{Institute of Industrial Science, The University of Tokyo, Meguro, Tokyo 153-8505, Japan}
\author{Y. J. Wang}
 \affiliation{Department of Physics, Northeastern University, Boston, Massachusetts 02115, USA}
\author{B. Barbiellini}
 \affiliation{Department of Physics, Northeastern University, Boston, Massachusetts 02115, USA}
\author{T. Ishikawa}
\affiliation{Institute of Space and Astronautical Science, Japan Aerospace Exploration Agency, Tsukuba, Ibaraki 305-8505, Japan}
\author{M. Itou}
\affiliation{Japan Synchrotron Radiation Research Institute, SPring-8 Sayo-cho, Hyogo 679-5198, Japan}
\author{Y. Sakurai}
\affiliation{Japan Synchrotron Radiation Research Institute, SPring-8 Sayo-cho, Hyogo 679-5198, Japan}
\author{A. Bansil}
\affiliation{Department of Physics, Northeastern University, Boston, Massachusetts 02115, USA}
\author{R. Ishikawa}
\affiliation{Institute of Industrial Science, The University of Tokyo, Meguro, Tokyo 153-8505, Japan}
\author{M. Hamaishi}
\affiliation{Institute of Industrial Science, The University of Tokyo, Meguro, Tokyo 153-8505, Japan}
\author{T. Masaki}
\affiliation{Department of Materials Science and Engineering, Shibaura Institute of Technology, Toyosu, Tokyo 135-8548, Japan}
\author{K. Kimura}
\affiliation{Department of Advanced Materials Science, The University of Tokyo, Kashiwa, Chiba 277-8561, Japan}
\author{T. Ishikawa}
\affiliation{RIKEN SPring-8 Center, Sayo-cho, Sayo-gun, Hyogo 679-5148, Japan}
\author{S. Nanao}
\affiliation{Institute of Industrial Science, The University of Tokyo, Meguro, Tokyo 153-8505, Japan}

\date{\today}

\begin{abstract}
Metallic liquid silicon at 1787K is investigated using x-ray Compton scattering. An excellent agreement is found between the measurements and the corresponding 
Car-Parrinello molecular dynamics simulations. Our results show persistence of covalent bonding in liquid silicon and provide support for the occurrence of theoretically 
predicted liquid-liquid phase transition in supercooled liquid states. The population of covalent bond pairs in liquid silicon is estimated 
to be 17{\%} via a maximally-localized Wannier function analysis. Compton scattering is shown to be a sensitive probe of bonding effects in the liquid state. 
\end{abstract}


\maketitle

Silicon (Si) presents a fascinating phase diagram as is the case in other systems that form tetrahedrally coordinated networks.\cite{kaczmarski} Upon melting, Si transforms into a 
metal accompanied by a density increase of about 10{\%}. The 
resistivity of liquid Si ($l$-Si) at the melting temperature T$_{m}$ is 0.75 
$\mu \Omega $m, which is comparable to that of simple liquid metals 
such as $l$-Al. However, the first neighbor atomic coordination number in 
$l$-Si remains 5.5$\sim $6~\cite{kim}, which is approximately 
half that of simple liquid metals, hinting that covalent bonds survive even in the metallic state~\cite{Chelikowsky}. In fact, molecular 
dynamics simulations of molten Si at 1800K suggest that approximately 30{\%} 
of the bonds are covalent and that these covalent bonds possess a highly 
dynamic nature, forming and breaking up rapidly on a time scale of 20 fs~\cite{stich}. It is remarkable that two completely different types of bonds$-$metallic and covalent$-$ can coexist in $l$-Si. 
In fact, the coexistence of two forms of liquid in a single component substance has been predicted to undergo a phase transition as a function of temperature and / or pressure~\cite{aptekar}, 
and many theoretical studies support the 
existence of a liquid-liquid phase transition (LLPT)~\cite{sastry, jakse, ganesh}. A recent study reports that $l$-Si could 
undergo an LLPT below about 1232K and above about -12kB, separating into a 
high-density metallic liquid (HDL) and a low-density semi-metallic liquid 
(LDL)~\cite{ganesh}. But, 1232K is far below the melting temperature of 1683K 
of Si, and as a result the supercooled state has remained inaccessible to current experimental techniques. Very recently, Beye \textit{et al.} 
have performed time-resolved x-ray measurements on Si using a femtosecond pulse-laser~\cite{beye} to reveal liquid polymorphs of Si which could support an LLPT, 
but these experimental conditions are far from being ideal~\cite{sastry, jakse, ganesh} so that the experimental confirmation of an LLPT in Si remains an open question.

A key requirement for the possibility of an LLPT obviously is that the metallic and covalent bonds coexist
in $l$-Si. Although experimental investigations of the atomic configuration 
hint at the existence of covalent bonds in $l$-Si, surprisingly, soft x-ray~\cite{hague} and magnetic susceptibility 
measurements~\cite{muller} of  
electronic properties so far do not support this viewpoint in that 
all four valence electrons in $l$-Si appear to behave like free-electrons. Emissivity and thermal conductivity of $l$-Si 
also exhibit a free-electron like temperature dependence~\cite{kobatake}. 
It is clear thus that the existence of the covalent bonds in $l$-Si is not well established. In this letter, Compton scattering experiments on Si 
are reported using a levitation technique. Experimental results are interpreted in terms of full quantum mechanical simulations to show the persistence of the covalent bond in $l$-Si. 
Moreover, by invoking a number of different first-principles approaches we show that Compton scattering spectroscopy is a sensitive probe of bonding effects 
even in the presence of disorder induced broadening of the spectrum in the liquid state.

As background, we note that Compton scattering refers to the inelastic x-ray scattering process at large energy and momentum transfers. To date, 
the technique has been applied primarily to investigate electronic structure and fermiology of solid-state systems, where the measured Compton profile is directly 
related to the momentum density distribution of the electronic ground state within the impulse approximation~\cite{IA,cooper04}. 
The present study shows how the Compton technique can also provide a novel spectroscopic window on the liquid state. 
Since no charged particles entering or leaving the sample are involved, Compton technique is a genuinely bulk probe, 
which is not complicated by surface effects present in photoemission or electron scattering experiments
\cite{arpes}.

Compton profiles of polycrystalline Si (300K) and molten Si (1787K) were 
measured by high-energy (116 keV) inelastic x-ray scattering at the BL08W beam line of SPring-8~\cite{hiraoka} with a momentum resolution of 0.16 a.u. 
The data processing to deduce the Compton profile from the raw energy 
spectrum consists of the following procedures: background subtraction and 
energy-dependent corrections for the Compton scattering cross-section, the 
absorption of incident and scattered x-rays in the sample, the efficiency of 
the analyzer and the detector, and correction for double scattering events; see Supplementary Material for details of data handling procedures. Backgrounds of the measured profiles are suppressed in the levitation 
technique used since no material is present in the immediate vicinity of the sample at the x-ray scattering center.

Molten Si is highly reactive with most crucibles. To hold the sample without 
contamination, a high temperature electrostatic levitator (HTESL) was 
used. The HTESL levitates a spheroid sample of 2mm diameter in a high-vacuum  environment (approximately 10$^{ - 5}$ Pa) using electrostatic forces via a 
feedback computer control~\cite{rhim93,masaki07}. Two separate Be windows (1.5 mm thick) were installed in the HTESL 
chamber for the entrance and exit of the x-ray beam. The scattering angle was set at 165 degrees. 
The sample (Si, 99.9999{\%} purity) was heated and melted using the focused radiation of three 50 W semiconductor laser beams emitting at 808 nm. Temperature 
was controlled within 15K and measured by pyrometry with wavelengths of 
0.90 and 0.96 nm.

Car-Parrinello Molecular Dynamics (CPMD) simulations~\cite{car} were performed with the Quantum-espresso package~\cite{giannozzi} within 
the framework of the density-functional theory using the  
generalized gradient approximation~\cite{perdew}. We employed 
norm-conserving pseudopotentials with plane-wave expansion of the Kohn-Sham 
wavefunctions and charge density up to a kinetic energy cutoff of 24 Ry and 96 Ry, respectively. 
Spin-unpolarized simulations were done with 64 Si atoms in simple cubic supercells. 
A time step of 0.097 fs was used with the 
fictitious mass of electron set to 300 a.u., which is within the suggested range for accurate CPMD simulations~\cite{grosmann, schwegler}. 
The solid Si ($s$-Si) simulation was performed at 300 K at the experimental density of 2.33 g/cm$^{3}$. 
The $l$-Si simulation was performed at 1787 K at the experimental density of 2.56 g/cm$^{3}$~\cite{rhim97}. Nose-Hoover thermostat~\cite{nose,hoover} on ions was 
applied in both solid and liquid simulations. We also used an additional 
Nose-Hoover thermostat on electrons in the liquid simulation due to the 
metallic character of the system. The solid and liquid simulations lasted 
for around 9 ps and 13.6 ps, respectively, after thermalization. The $s$-Si 
Compton profile was calculated from the average of ten uncorrelated 
configurations, equally spaced by 1 ps. The $l$-Si Compton profile was 
calculated from the average of ten uncorrelated configurations, equally 
spaced by 1.5 ps. Both profiles were spherically averaged and convoluted 
with a Gaussian to match the 0.16 a.u. experimental resolution. 
In order to estimate disorder effects in the liquid \cite{disorder_effect}, we have performed the simulation 
on $l$-Si using the first-principles Korringa-Kohn-Rostoker coherent potential approximation (KKR-CPA) 
framework \cite{bansil}.
In these simulations, we adopted the lattice model for $l$-Si given in Ref.~\cite{sahara}. 
In particular, we used a body centered cubic structure with lattice constant 5.13 a.u. in which lattice sites 
are occupied by randomly distributed 50{\%} Si atoms and 50{\%} empty spheres.

\begin{figure}
 \includegraphics[width=9cm,clip]{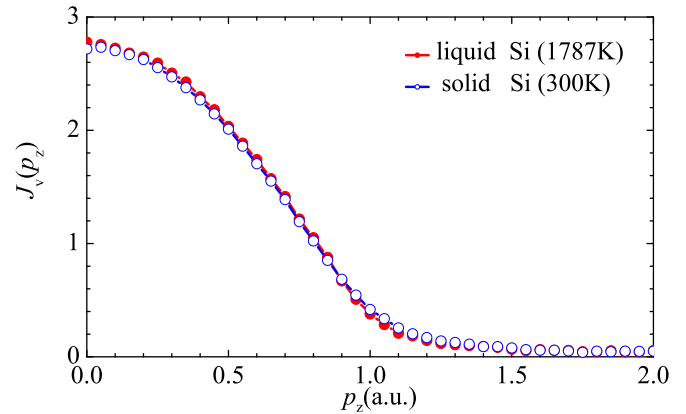}
 \caption{\label{Fig. 1}   Experimental valence-electron Compton profile of solid (blue open circles) and liquid Si (red solid circles). 
 The error bars are less than the symbol size.}
\end{figure}

The valence-electron Compton profiles of $s$-Si (300K) and $l$-Si (1787K), obtained by subtracting the theoretical core-electron 
profile $J_{c}(p_{z})$ from the experimental profiles, are presented in Fig. 1. The theoretical core-electron profiles 
are based on the free-atom Hartree--Fock simulations~\cite{biggs}, 
where (1s)$^{2}$(2s)$^{2}$(2p)$^{6}$ are treated as core electrons. 

Figure 2 presents the differences between the Compton profiles for the solid and liquid phases 
[$\Delta J(p_z)=J_{solid}(p_{z}) - J_{liquid}(p_{z})$]. 
The results of CPMD simulation are obtained by taking the difference of the liquid Compton 
profile and the spherically averaged profile of the solid. The CPMD results 
agree very well with the experimental results, establishing the 
efficacy of the bonding properties of $l$-Si obtained through CPMD simulation~\cite{stich}. 
The difference between the KKR-CPA results for $l$-Si 
and the spherically averaged $s$-Si is also shown in Fig. 2. 
The effect of disorder in the KKR-CPA calculation yields a momentum density 
smearing near $p_z=1$ a.u. \cite{bansil} and
produces a dip of $\Delta J(p_z)$ at this same location.
However, in the experiment $\Delta J(p_z)$ shows a peak near $p_z=1$ a.u. instead of a dip.
We therefore conclude that the features induced by the Si site disorder are overshadowed 
by competing effects due to the covalent bond breaking. 

\begin{figure}
 \includegraphics[width=8.8cm,clip]{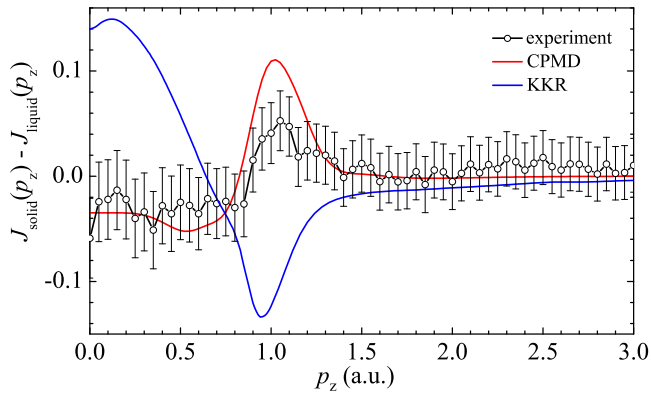}
 \caption{\label{Fig. 2} Differences between the Compton profiles for the solid and liquid phases, 
[$J_{solid}$ ($p_{z})$ -- $J_{liquid}$ ($p_{z})$]: experiment (black), CPMD simulation (red), and KKR-CPA simulation (blue). }
\end{figure}

In order to analyze the bonding character in $l$-Si obtained through CPMD 
simulation, the maximally-localized Wannier functions (MLWF) 
analysis method is used~\cite{marzari}. MLWF's are constructed through unitary 
transformation of Kohn-Sham wavefunctions such that the average spread 
(variance) of the Wannier functions is minimized. Same as the Kohn-Sham 
wavefunctions, each MLWF represents two electrons in spin-unpolarized 
calculations. In the $s$-Si simulation, the spreads 
of MLWF's are all smaller than 1.84 {\AA}$^{2}$ and all the corresponding 
Wannier centers (WC) are in the middle between two Si atoms, showing that 
all electron pairs in $s$-Si are covalent. On the other hand, MLWF's calculated 
from the $l$-Si simulation exhibit diverse characteristics. In particular, a 
majority of MLWF's have spreads larger than 1.84 {\AA}$^{2}$, up to over 7.8 
{\AA}$^{2}$. The spread is an indicator of the spatial extent of the MLWF. 
The inability of the MLWF procedure to localize some of the Wannier functions is 
expected as the system is metallic and the MLWF's with large spreads contribute 
to the metallic properties of  $l$-Si. We refer to such electron pairs as diffuse pairs. 
The MLWF's with small spreads mostly correspond to the covalent electron pairs in $l$-Si. 
However, there is a special type of MLWF whose spread is comparable to that of the covalent electron pairs 
but the WC is not shared by any two Si atoms, so that the corresponding electron pair may be called a lone 
pair. The lone pair MLWF is computed to have a spread of 2.21 {\AA}$^{2}$~\cite{Patrick}.

In order to categorize different types of electron pairs in $l$-Si,
we sort the MLWF's into three groups: 
covalent bond pairs, lone 
pairs, and diffuse pairs. All electron pairs with 
Wannier spread larger than 2.21 {\AA}$^{2 }$ are assigned to be diffuse 
pairs. Among electron pairs with spreads smaller than 2.21 {\AA}$^{2}$, an 
electron pair is a covalent bond pair if the following two criteria are 
satisfied: (1) The WC lies between two Si atoms separated by less than 3.1 
{\AA}, which corresponds to the first minimum of the Si-Si radial 
distribution function; (2) The WC lies within 1.36 {\AA} from the mid-point 
of the two Si atoms. The 1.36 {\AA} equals square root of 1.84 
{\AA}$^{2}$,$^{ }$ which is the largest variance of covalent bond pair 
MLWF's in $s$-Si. With this scheme, we found that on the average, 
there are 17{\%}, 83{\%}, and less than 1{\%} of covalent bond pairs, 
diffuse pairs and lone pairs, respectively~\cite{footnote}. Figure 3 shows 
a snapshot from the simulation carried out on $l$-Si. 
Note that the population of different electron pair types depends on the criteria adopted. 
The covalent bond pair population of 17{\%} can be taken as the lower-bound. 
This is because the choice of the Wannier spread cutoff of 2.21 {\AA}$^{2}$ is too tight for a
bond pair or a lone pair since MLWF's are expected to be more dispersed in 
the metallic liquid phase. 

The key point of our analysis is that the existence of covalent bonds in metallic $l$-Si 
is clearly confirmed, and that liquid Si is not homogeneous at 
the atomic scale. The coexistence of two different bonding natures in a single 
component liquid is a precondition that deeply 
supercooled $l$-Si can undergo LLPT~\cite{aptekar}. Our study thus support 
the possible occurrence of LLPT. 

The temperature of $l$-Si in our study is 1787K which is far above the 
predicted LLPT occurrence temperature of 1232K~\cite{ganesh}. To infer what 
kind of change in $l$-Si could occur and cause LLPT with decreasing 
temperature, it is useful to look at the temperature dependence of density 
of $l$-Si. The density of $l$-Si has been measured between 1350K and 1850K~\cite{rhim97}; a quadratic behavior of 
the density as a function of temperature in the supercooled region has been observed. 
This is interpreted as the coexistence of covalent and metallic bonds~\cite{tanaka}. Above the melting point, 
the temperature dependence of the density shows approximate linearity, indicating little heterogeneity in bonding properties. 
In fact, our study shows that the majority of the electron pairs are in metallic bonds. In supercooled states, the ratio of covalent bonds 
could gradually increase with decreasing temperature since the temperature dependence of density deviates from linearity and becomes quadratic. 
The covalent bond pairs could assemble to form LDL domains. When temperature drops to 1232K where LLPT could take place, the domain size 
could exceed a critical size as in the case of nucleation in liquids, separating $l$-Si into HDL and LDL.

\begin{figure}
 \includegraphics[width=9cm,clip]{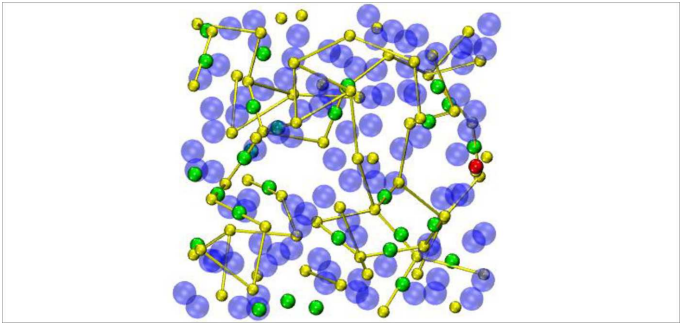}
 \caption{\label{Fig. 3}   Snapshot from the simulation on $l$-Si at 1787K exhibiting Si atoms (yellow), covalent bond pairs (green), lone pairs (red), 
and diffuse pairs (translucent blue). Note, bonds connecting Si atoms are only guides for the eye as they do not correspond to the chosen cutoff bond length of 3.1 {\AA}.}
\end{figure}

Finally, as stated in the introduction, there has been a longstanding controversy related to the interpretation of the nature of bonding in $l$-Si: 
atomic structure experiments have given indications of presence of covalent bonds~\cite{Chelikowsky,covalent}, 
while other experiments suggest that all valence electrons behave like free-electrons~\cite{hague,muller,kobatake}. 
Our study resolves this controversy by establishing the persistence of a significant number of covalent bonds in $l$-Si, even though a majority of valence electrons in $l$-Si exist as diffuse pairs.

To summarize, we have carried out Compton scattering measurements on $s$- and 
$l$-Si coupled with CPMD and KKR-CPA simulations. Comparison with the 
KKR-CPA results shows that Compton scattering remains a sensitive probe of bonding properties despite the smearing of the momentum density induced by disorder in the liquid state. 
The excellent agreement of the CPMD simulation with the experimental results establishes the existence of covalent bonds in metallic $l$-Si. The contributions of the covalent bond 
pairs, diffuse pairs, and lone pairs are estimated to be about 17 {\%}, 
83{\%} and less than 1{\%}, respectively. Our study thus provides evidence for 
the coexistence of two distinct bonding species, metallic and covalent, which is a prerequisite for the occurrence of LLPT in deeply 
supercooled $l$-Si, and thus supports the possibility of such a phase transition. Future Compton experiments with intense x-ray pulses will 
be needed to probe how the silicon bonding properties evolve from metallic to covalent character and how an LLPT may occur.

We acknowledge useful discussions with S. Kaprzyk.
The Compton profile measurements were performed with the approval of JASRI 
(Proposal Nos. 2005B0317, 2006A1389, 2007B1235). The work at JAXA was 
supported by the Sumitomo Foundation and Grants-in-Aid for Scientific 
Research KAKENHI from MEXT of Japan under Contract Nos. 16206062 and 
20760504. The work was supported by the US Department of Energy, Office of Science, 
Basic Energy Sciences grants DE-FG02-06ER46344 at Princeton University, and DE-FG02- 07ER46352 and 
DE-FG02- 08ER46540 (CMSN) at Northeastern University, 
and benefited from allocation of time at NERSC and NU's Advanced Scientific 
Computation Center.


\begin{thebibliography}{}
\bibitem{disorder_effect}
The effect of disorder on momentum density is very different from the temperature dependence 
of the Compton profile in solids, the latter being mainly dominated by the variation of the lattice constant.
In silicon, the variation of the lattice constant with temperature is less than 1{\%}, which yields a negligible effect on the Compton profile.
\bibitem{Patrick}
We calculated the MLWF's of an isolated [Si$_{4}$H$_{9}$] cluster, 
which is a simple model for a Si-based system containing a lone pair. In this 
cluster, a central Si atom is 3-fold coordinated by Si atoms, which are in 
turn saturated by H atoms, so that an electron lone pair resides on the 
central Si atom.
\end{thebibliography}
\end{document}